\documentclass[reprint,
amsmath,
amssymb,
aps,
superscriptaddress, 
prapplied,
longbibliography
]{revtex4-2}

\usepackage{graphicx}
\usepackage{dcolumn}
\usepackage{bm}
\usepackage{amsmath,amssymb}
\usepackage{url}
\usepackage{color}
\usepackage{hyperref}
\usepackage{tikz}
\usetikzlibrary{arrows.meta, positioning, calc, fit}

\begin{document}

\preprint{APS/123-QED}

\title{Hierarchical QAOA for the Vehicle Routing Problem via Clustered Decomposition and Local Feasibility Repair}

\author{Shreetam Dash}
\email{shreetamdash.16@gmail.com}
\affiliation{Center for Quantum Science and Technology, Siksha 'O' Anusandhan (Deemed to be University), Khandagiri Square, Bhubaneswar-751030, Odisha, India}

\author{Shreya Banerjee}
\altaffiliation[s.banerjee3@exeter.ac.uk]{}
\affiliation{Center for Quantum Science and Technology, Siksha 'O' Anusandhan (Deemed to be University), Khandagiri Square, Bhubaneswar-751030, Odisha, India}
\affiliation{Department of Physics and Astronomy, University of Exeter, Stocker Road, Exeter-EX4 4QL, United Kingdom}

\author{Prasanta K. Panigrahi}
\altaffiliation[director.cqst@soa.ac.in]{}
\affiliation{Center for Quantum Science and Technology, Siksha 'O' Anusandhan (Deemed to be University), Khandagiri Square, Bhubaneswar-751030, Odisha, India}
\affiliation{Department of Physical Sciences, Indian Institute of Science Education and Research Kolkata, \\ Mohanpur- 741246, West Bengal, India}

\begin{abstract}
We propose a hierarchical quantum approximate optimization framework for solving large-scale Vehicle Routing Problems (VRP) using Quantum Approximate Optimization Algorithm (QAOA). The method decomposes a VRP instance into balanced clusters of customer nodes. We formulate intra-cluster routing as Open loop Traveling Salesman Problems (OTSPs), and inter-cluster routing as a reduced VRP over the cluster representatives and depot. We then map the sub-problems to Ising Hamiltonians and solve with both standard and multi-angle QAOA variants at fixed depth $p=3$, and merge them to produce a routing path for the original VRP. Additionally, to improve solution feasibility and success probability, we introduce a polynomial-time post-processing protocol that samples candidate bit-strings from the QAOA output using a probability threshold and performs exhaustive local $1$ and $2$ bit-flip searches around these candidates. Benchmarking on 100 randomly generated 13-node, two-vehicle VRP instances, we show that the post-processed standard-QAOA implementation achieves high success rates and approximation ratios within 1.2–1.5 compared to classical optimizer (Gurobi) solutions, while requiring only 12 logical qubits per subproblem instead of 156 qubits for a direct edge-based encoding. These results provide a proof-of-concept demonstration that hierarchical decomposition, shallow QAOA, and local bit-flip repair can offer a scalable and resource-efficient pathway toward larger VRP instances on near-term quantum devices.
\end{abstract}

\maketitle
\section{Introduction}\label{sec:introduction}
The Vehicle Routing Problem (VRP) is one of the fundamental and challenging combinatorial optimization problems with important large-scale, practical applications in logistics, supply chain management, and transportation. The problem is used to determine the optimal routes for a fleet of vehicles to serve a set of customers while minimizing total energy costs~\cite{Azad2022,Fitzek2024,Kerscher2024b}. As a generalization of the Traveling Salesman Problem (TSP), VRP belongs to the class of NP-hard problems ~\cite{doi:10.1137/1.9780898718515.ch1,somvanshi2026quantum,logistics10010013}, which makes finding exact solutions extremely difficult for large-scale instances~\cite{maciejunes2025solvinglargescalevehiclerouting}.

Recent advances in quantum computing have opened new passages for solving these computationally challenging optimization problems. The Quantum Approximate Optimization Algorithm (QAOA), introduced in \cite{Farhi2014}, has emerged as a promising hybrid quantum-classical approach for solving combinatorial optimization problems on near-term quantum devices~\cite{Azad2022,Drapeau_2026,s5jv-jh24, p2lg-z4kn}. However, the scalability limitations of current quantum hardware have restricted practical implementations to small-scale instances, with most existing studies limited to problems involving 4-6 locations and 2-3 vehicles~\cite{Azad2022,Huang2025, perlin2026faulttolerantexecutionerrorcorrectedquantum, Irie2019, feld2019hybrid}.

To overcome the scalability challenges in quantum optimization, decomposition strategies have proven invaluable in the classical VRP domain. In Clustering-based decomposition methods~\cite{Patil2024,morampudi2017clustering}, those who implement K-means algorithms, have got significant success in converting large-scale VRP instances into smaller manageable subproblems~\cite{Alfiyatin2018,Borowski2020}. These cluster-first, route-second approaches allow one to apply optimization techniques to problems that would otherwise be computationally challenging~\cite{Alesiani2022,Kerscher2024}.
The effectiveness of clustering lies in its ability to reduce the complexity of the problem while maintaining the quality of the solution~cite{}. By grouping the customers into clusters, the overall problem can be decomposed into intra-cluster routing (typically formulated as Open loop TSP instances) and inter-cluster routing (maintaining the VRP structure)~\cite{Santini2023,Kerscher2024}. This  decomposition has enabled classical algorithms to handle instances with large number of customers while maintaining near-optimal solution quality~\cite{Kerscher2024,DondoCerda2007, Padmasola2025}.

The implementation of QAOA \cite{Farhi2014,svensson2023hybrid} and its many variants \cite{Herrman2022,9259965, 10.1145/3549554, le2023quantum} in large-scale optimization problems encounters limitations in circuit depth and parameter optimization. Alternate layers of the cost and mixer Hamiltonians improve the performance of the algorithm with increasing circuit depth (parameter $p$~\cite{Farhi2014, a12020034,Kotil2025}. However, circuits with higher depth are more open to noise and decoherence in current quantum hardware, creating a fundamental trade-off between solution quality and practical implementation~\cite{Pellow-Jarman2024}.

This work adapts the hierarchical approach to solve large-scale VRP, and show that QAOA with a fixed depth ($p=3$) can be useful for practical applications of combinatorial optimization. Our method demonstrates an effective path to utilize quantum approximate optimization for large-scale Vehicle Routing Problems on near-term quantum computers with a handful of logical qubits \cite{perlin2026faulttolerantexecutionerrorcorrectedquantum}. Additionally, we propose a local bit-flip search post-processing scheme that aids QAOA to achieve high success rates and near-ideal approximation ratios, and show that our method is robust against low clusterability \cite{lawson1990hopkins,1375706} of a dataset. 

This article is arranged as follows. In Sec.~\ref{sec:methods}, we present a comprehensive mathematical formulation of the decomposition of a large-scale VRP into smaller subproblems (Open loop TSP on clusters of customer nodes, and a VRP for inter-cluster routing). We implement the standard QAOA \cite{Farhi2014}, and its Multi-Angle variant \cite{Herrman2022} to solve these problems, and benchmark our methodology in \ref{sec:results}. Finally we conclude in sec.~\ref{sec:conclusion}. 

\section{Methodology} \label{sec:methods} 

\begin{figure*}[t]
    \centering
    \resizebox{0.9\textwidth}{!}{%
    \begin{tikzpicture}
        \node[anchor=south west, inner sep=0] (main) at (-50,0)
        {\includegraphics[width=0.60\linewidth]{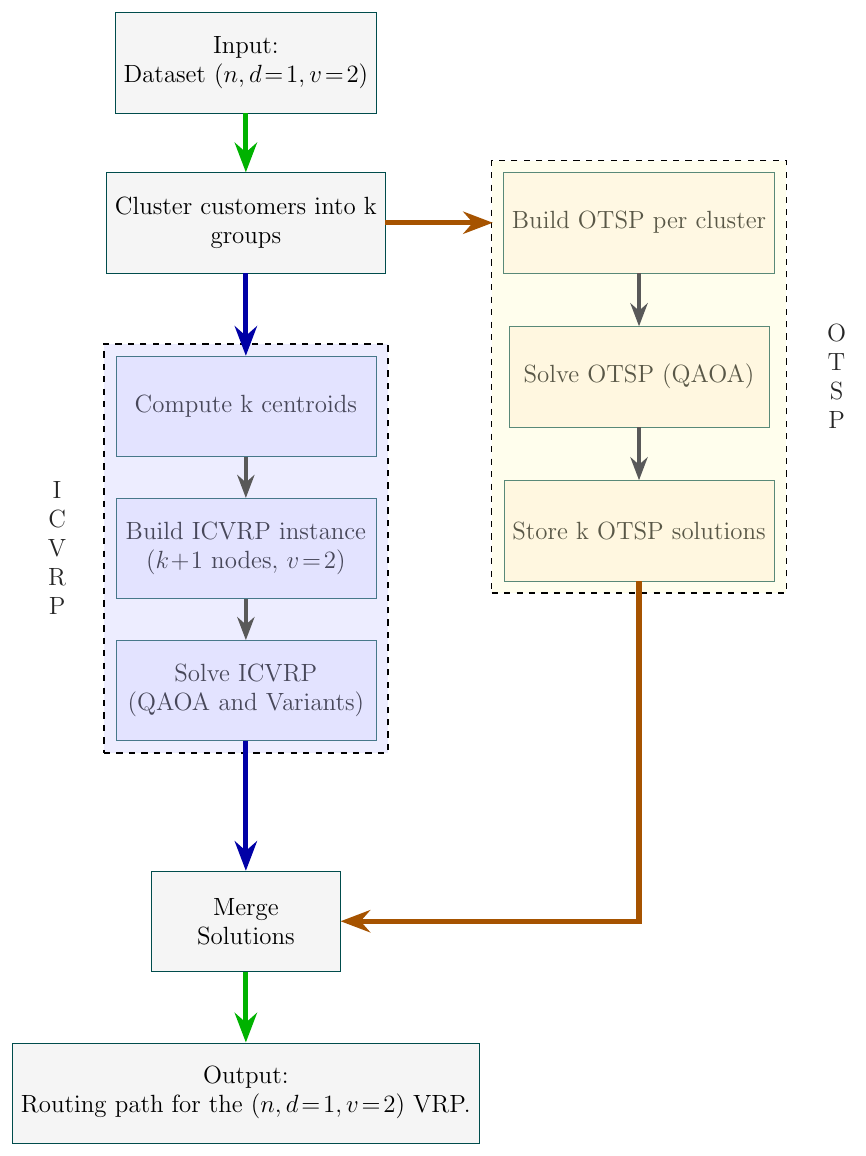}};

        \node[anchor=north east, inner sep=0] (inset) 
        at ([xshift=0.4\linewidth,yshift=0.00\linewidth]main.north east)
        {\includegraphics[width=0.5\linewidth]{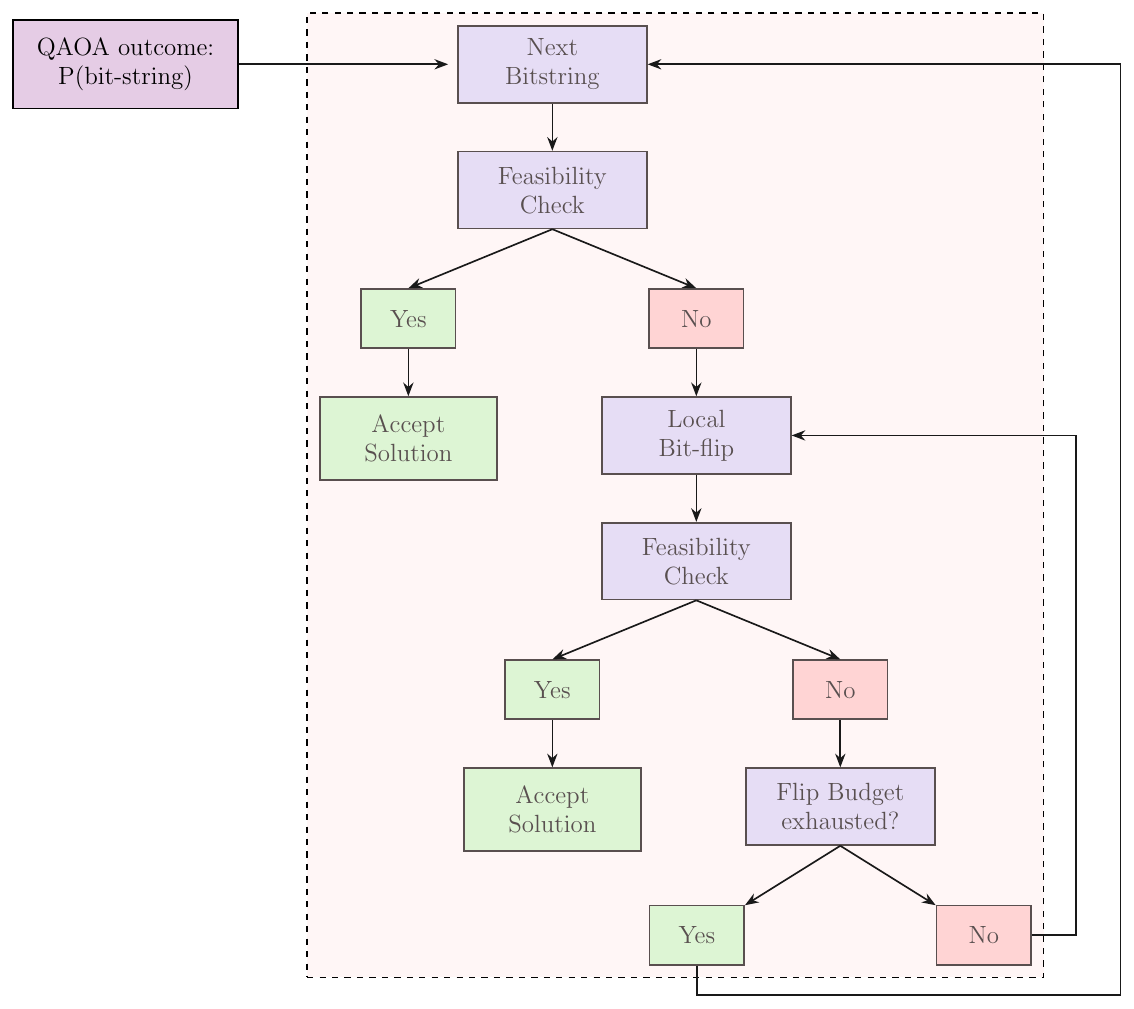}};

        \node[anchor=north west, font=\small] 
        at ([xshift=2pt,yshift=-2pt]main.north west) {(a)};

        \node[anchor=north west, font=\small] 
        at ([xshift=-16pt,yshift=-2pt]inset.north west) {(b)};

        \node[anchor=south east, inner sep=0] (inset2) 
        at ([xshift=0.4\linewidth,yshift= -0.05\linewidth]main.south east)
        {\includegraphics[width=0.4\linewidth]{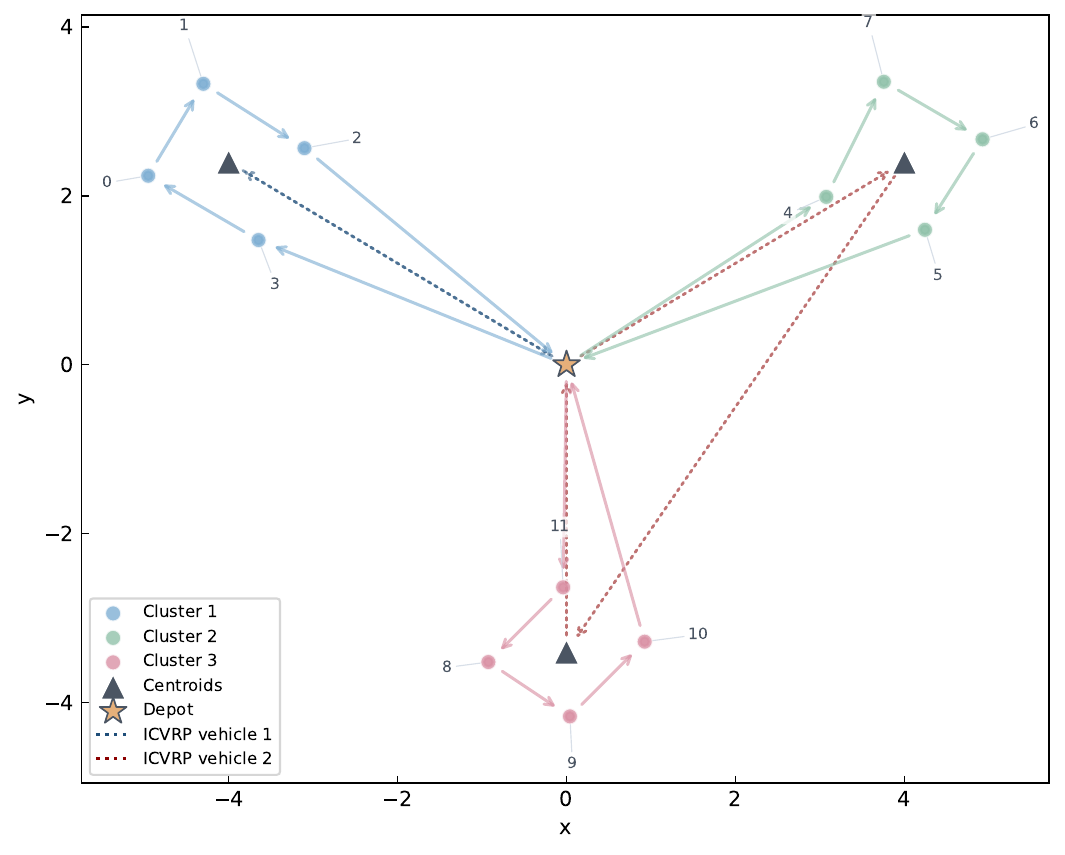}};

        \node[anchor=north west, font=\small] 
        at ([xshift=-16pt,yshift=-2pt]inset2.north west) {(c)};
    \end{tikzpicture}%
    }

    \caption{\textbf{(a)}: Workflow of the proposed hierarchical quantum approximate algorithm to solve large-scale VRP. An input dataset with $n$ customers is divided into $k$ clusters (gray block). For each cluster, an Open loop Traveling Salesman Problem (OTSP) is then formulated and solved (yellow blocks). Centroids of each cluster are computed as cluster representative nodes, and Inter Cluster Vehicle Routing Problem (ICVRP) is formulated and solved over the depot and centroids (blue blocks). Next, solutions of the subproblems are merged, and a routing path for the original VRP is produced. \textbf{(b)}: Post-processing algorithm to increase the approximation ratio of QAOA solutions and success rates. Based on a user-defined threshold, a polynomial (in number of qubits) number of solutions from the QAOA output is selected as candidate bit-string list. The bit-strings are then checked to determine whether they satisfy the problem constraints. If not, an exhaustive $1$ and $2$ local bit-flip search is employed to check whether a feasible solution can be found. \textbf{(c)} Merged outcome of the original VRP, shown with an example dataset. The depot is shown with a yellow star, and customer nodes divided into three clusters are marked in blue, green and pink. Directed solid arrows of the same colors mark the OTSP routing paths. Centroids of each cluster are marked as black triangles. The ICVRP routing paths for vehicles $1$ and $2$ are respectively depicted with directed dashed blue and red arrows.}
    \label{fig:methodology}
\end{figure*}

As mentioned in Sec.~\ref{sec:introduction}, we decompose a large-scale VRP by clustering the customer nodes into smaller datasets. The routing problem within each cluster is then effectively formulated as an Open loop Traveling Salesman Problem (OTSP), where vehicles are not required to return to their initial point within the cluster. This process is beneficial in the context of multi-cluster VRP, as it allows for optimal transitions between clusters, maintaining computational efficiency~\cite{Padmasola2025}. 

Once the routing path within each cluster is formulated, an inter-cluster routing problem is formulated, maintaining actual VRP structure, where vehicles optimize the routes between clusters and the depot. This approach thus reduces the size of the effective problems to be solved, with a simple trade of increasing the number of problems to be solved. 

This section presents a systematic approach to solve the large-scale Vehicle Routing Problems using quantum optimization algorithms. We propose a three-step approach to approximately solve a large-scale Vehicle Routing Problem using quantum computers with a handful number of logical qubits. First we decompose the VRP into multiple smaller optimization problems using a distance (weight)-based clustering of nodes. Further, we consider each cluster as a single node through a representative point, and formulate and solve a final routing problem. Finally, we combine the solutions, providing a scalable, approximate solution of the original NP-hard problem. The approach is outlined in Fig.~\ref{fig:methodology}(a).

\noindent\textbf{Clustering of Nodes}: To efficiently cluster the $n$ customer node into $k$ clusters, we employ a popular classical unsupervised clustering algorithm, i.e., K-means algorithm \cite{Alfiyatin2018}. To ensure a balanced workload distribution, the clustering objective minimizes the sum of squared distances within the clusters, i.e., 
\begin{equation*} 
\min \sum_{i=1}^{k} \sum_{j \in C_i} \|x_j - \mu_i\|^2 
\end{equation*} 
where \(C_i\) represents cluster \(i\), \(x_j\) denotes the coordinates of node \(j\), and \(\mu_i\) is the centroid of cluster \(i\).  
 
After the unsupervised clustering is performed, for each cluster, we assign its centroid as its representative node. The clusters are then considered as a singular node when optimizing the inter-cluster transfers. The choice of centroid ensures the fairness in assigning travel cost while solving the inter cluster VRP. This choice, alongwith specific penalty functions (see below) ensure that vehicles avoid unnecessary detours including purposelessly returning to their starting points. Further, this approach minimizes travel costs by guiding vehicles directly toward the next cluster or back to the depot from the representative node following the optimized path. 

Throughout Sec.~\ref{sec:methods}, we illustrate our methodology with a concrete example using a $13$-location VRP instance consisting of $12$ customer nodes and $1$ depot, served by $2$ vehicles. For simplicity, we only consider distance as the weights associated with each possible route. The example dataset of $12$ customer nodes and $1$ depot with their coordinates is provided in Eq.~\ref{eq_dataset1}.

\begin{widetext}
\begin{equation}\label{eq_dataset1}
\begin{aligned}
\text{Depot:} \; D: [50.00, 50.00], \\
\text{Customers:} \quad 1: [72.49, 84.01], &\; 2: [79.64, 76.97], \; 3: [68.12, 68.12], \; 4: [66.16, 82.32],\\
5: [77.02, 29.16], &\; 6: [65.41, 34.40], \; 7: [81.65, 19.25], \; 8: [68.64, 18.67],\\
9: [21.08, 25.50], &\; 10: [23.64, 20.82], \; 11: [27.24, 17.79], \; 12: [20.84, 22.33]
\end{aligned}
\end{equation}
\end{widetext}

After clustering the example dataset, distance matrices are constructed for each cluster based on Euclidean distances between the nodes. For the three clusters obtained from the example dataset, the distance matrices are:

\noindent \textit{Cluster 1:}
\begin{equation*}
W_1 = \begin{pmatrix}
0 & 10.03 & 16.48 & 6.55 \\
10.03 & 0 & 14.53 & 14.50 \\
16.48 & 14.53 & 0 & 14.33 \\
6.55 & 14.50 & 14.33 & 0
\end{pmatrix}
\end{equation*}

\noindent \textit{Cluster 2:}
\begin{equation*}
W_2 = \begin{pmatrix}
0 & 12.74 & 10.94 & 13.43 \\
12.74 & 0 & 22.21 & 16.06 \\
10.94 & 22.21 & 0 & 13.02 \\
13.43 & 16.06 & 13.02 & 0
\end{pmatrix}
\end{equation*}

\noindent \textit{Cluster 3:}
\begin{equation*}
W_3 = \begin{pmatrix}
0 & 5.33 & 9.87 & 3.18 \\
5.33 & 0 & 4.71 & 3.18 \\
9.87 & 4.71 & 0 & 7.85 \\
3.18 & 3.18 & 7.85 & 0
\end{pmatrix}
\end{equation*}
 
\noindent\textbf{Intra-cluster Routing: Open Loop TSP}: As can be seen from Fig.\ref{fig:methodology}(a), after clustering, we formulate an intra-cluster routing problem as an Open Loop Traveling Salesman Problem (OTSP) to optimize vehicle paths within every cluster. This choice is made  as the vehicles only need to enter and leave each cluster just once. The OTSP guides the vehicle along an optimally directed path from the initial node to the final inside a cluster, while properly satisfying all routing constraints. 

For a cluster with nodes \(\{1, 2,....n\}\), the OTSP is formulated using $n(n-1)$ binary decision variables \(x_{ij} \in \{0,1\}\), where \(x_{ij} = 1\) if there is an existing route from node \(i\) to node \(j\).

To solve the OTSP for each cluster, we construct an objective function that minimizes the total travel distance for one particular cluster. We consider the first cluster \(\{1, 2,3,4\}\) as an example. As mentioned earlier, $x_{ij}$ represents a binary decision variable that determines the existence of a route between node $i$ and $j$. The objective function for OTSP within Cluster $1$ can then be written as, 

\begin{equation}\label{ostp_obj}
H^{1}_{OTSP}= \min \left( \sum_{i=1}^{4} \sum_{j=1}^{4} w_{ij} x_{ij} \right)
\end{equation} 
where \(w_{ij}\) represents the Euclidean distance between nodes \(i\) and \(j\), i.e., the elements of matrix $W_1$.

\vspace{2 mm} 
The OTSP formulation includes considering several constraint categories for valid routing solutions, combined with the objective function. Below we consider the constraints pertaining to our specific problem and add them to the objective function as penalty terms that violates the constraints. This way, one can formulate a minimizable objective function that follows the constraints alongwith the main objective function.

For the selection of initial and final nodes for open loop TSP within each cluster, we adopt a systematic approach based on minimizing distances to both the depot and neighboring clusters. The initial point is designated as the node within each cluster that exhibits the minimum combined distance to the VRP depot and to representative nodes in other clusters, while the final point is identified as the node with the second-minimum combined distance. This strategic selection ensures that both points remain closely aligned with each other. The motivation behind selecting two nearby points is to replicate the closed-loop nature of traditional TSP formulations, where vehicles are required to return to the initial point thereby establishing a direct link between the final and initial locations.
Further, in our example data set for the 1st cluster, we have calculated nodes 3 and 4 for the initial and final representative points respectively.

\begin{itemize}
\item{\textit{Outgoing Edge Constraints}: Excluding the final location (in our example, node $3$), each node must have exactly one outgoing edge. Mathematically, this constraint can be expressed as,  
\begin{equation*} 
\sum_{j \neq i} x_{ij} = 1, \quad \forall i \in \{1, 2, 4\}.
\end{equation*} 
One can write the corresponding penalty term as a quadratic minimizing term as,  
\begin{equation} 
H^{P}_{Out} = \sum_{i \in \{1,2,4\}} \left(1 - \sum_{j \neq i} x_{ij}\right)^2 
\end{equation}
}

\item {\textit{Incoming Edge Constraints}: Excluding the initial location (in our example, node $4$), each node must have exactly one incoming edge, i.e, 
\begin{equation*} 
\sum_{i \neq j} x_{ij} = 1, \quad \forall j \in \{1, 2, 3\}.
\end{equation*}

Similar to the previous case, this constraint can also be written as an optimizing Hamiltonian,  
\begin{equation} 
H^{P}_{In} = \sum_{i \in \{1,2,3\}} \left(1 - \sum_{j \neq i} x_{ji}\right)^2.
\end{equation} 
}

\item {\textit{Sub-tour Elimination}: Additional constraints are introduced to penalize invalid subtours: 
\begin{equation*} 
x_{12} + x_{24} + x_{41} \leq 2, 
\end{equation*} 
which leads to a penalty term as, 
\begin{equation}
H^{P}_{SE} = \left(2 - \left(x_{12} + x_{24} + x_{41}\right)\right)^2.
\end{equation}
}

\item{\textit{Incoming and Outgoing edges from Initial and Final nodes:}} Traditionally, one also considers constraints for selection of initial and final Nodes. As the initial node has no incoming edges, and the final node has no outgoing edges, they can be expressed as,  
\begin{equation}\label{in-out}
\sum_{j} x_{j,\text{initial}} = 0, \quad \sum_{j} x_{\text{final},j} = 0.
\end{equation} 

\noindent As explained earlier, we deterministically select the initial and final nodes to closely replicate the traditional TSP formulation. The corresponding terms to Eq.~\ref{in-out} is thus eliminated from the expression of $H^{1}_{OTSP}$ to satisfy the constraints.  
\end{itemize}

Finally, the Open loop Traveling Salesman Problem is formulated as a Quadratic Unconstrained Binary Optimization (QUBO) problem suitable for quantum optimization for each cluster. The complete QUBO Hamiltonian combines the objective functions $H^{1}_{OTSP}$ and penalty terms $H^{P}_{Out}, H^{P}_{In}$, and $H^P_{SE}$ as: 
\begin{equation} 
H_{\text{OTSP}} = \sum_{i,j} w_{ij} x_{ij} + A (H^{P}_{In} + H^{P}_{out}+ H^{P}_{SE}), 
\end{equation} 
where $i \in \{1, 2, 3\}$ and $j \in \{2, 3, 4\}$. \(A = 50\) is a penalty parameter, chosen to be greater than the maximum weight between any pair of nodes. 

Expressing the cost Hamiltonian $H_{OTSP}$ in its complete QUBO formulation, one can easily map it to its Ising version for QAOA implementation \cite{qubo_ising,PhysRevApplied.22.064068}. The QUBO expression for a quadratic cost function $f(x)$ is given as, 

\begin{equation*} 
f(x)_{\text{QUBO}} = \vec{x}^T Q \vec{x} + \vec{g}^T \vec{x} + c
\end{equation*} 
where \(Q\) is the quadratic coefficient matrix, \(\vec{g}\) contains linear terms, and \(c\) is a constant offset. 

The mapping of a QUBO cost function to an Ising Hamiltonian uses the transformation \(x_{ij} = (s_{ij} + 1)/2\), where \(s_{ij} \in \{-1, 1\}\) are spin variables. The corresponding Ising Hamiltonian is given as, 
\begin{equation*} 
H_{\text{Ising}} = -\sum_{i} \sum_{j<i} I_{ij} s_i s_j + \sum_{i} h_i s_i + d, 
\end{equation*} 
where,  

\begin{align} \label{Ising}
\nonumber I_{ij} &= -\frac{Q_{ij}}{4} \\ 
\nonumber h_i &= \frac{g_i}{2} + \sum_j \frac{Q_{ij}}{4} + \sum_j \frac{Q_{ji}}{4} \\ 
d &= c + \sum_i \frac{g_i}{2} + \sum_{i,j} \frac{Q_{ij}}{4} 
\end{align} 
Following the same recipe as Eq.~\ref{Ising}, we have decomposed the OTSP cost function $H_{OTSP}$ into single-spin, and two-spin interaction terms, as well as a constant. In a QAOA formulation \cite{Farhi2014}, this cost function reads, 

\begin{align}\label{qaoa_cost}
H_{\text{cost}} &= -\sum_{i<j} I_{ij} \sigma_i^z \sigma_j^z - \sum_i h_i \sigma_i^z - d.
\end{align}

\noindent\textbf{Inter-cluster VRP}: To approximately solve the Vehicle Routing Problem on all nodes, we now need to address the assignment of vehicles by considering individual clusters as single locations, and optimize for the optimal route between the clusters. In this subsection, we present a concrete formulation for the reduced VRP prior using Multi-Angle QAOA (MA-QAOA) \cite{Herrman2022} to find an optimal solution for the problem.

For each cluster \(C_i\) (\(i = 1, 2, 3\)), we compute its centroid, which serves as the representative node of $C_i$ for the inter-cluster routing. The centroid effectively and efficiently serves as a connection point for inter-cluster routing while maintaining geometric consistency.

Next, inter-cluster distances are computed between the representative points of all clusters, as well as to and from the depot, and an inter-cluster distance matrix  is constructed. This distance matrix provides a reduced geometric approximation of the overall Vehicle Routing Problem. 
For the example dataset illustrated above, the inter-cluster distance matrix is: 

\begin{equation*} 
W_C = \begin{pmatrix} 
0 & 33.83 & 35.25 & 39.04 \\ 
33.83 & 0 & 52.51 & 50.13 \\ 
35.25 & 52.51 & 0 & 74.21 \\ 
39.04 & 50.13 & 74.21 & 0 
\end{pmatrix}.
\end{equation*} 
 
The original Vehicle Routing Problem with $12$ customer nodes, $1$ depot, and $2$ vehicles can now be formulated as a clustered Vehicle Routing Problem with $2$ vehicles serving \textit{only $3$} customer locations, and $1$ depot, where the clustered customer locations are represented by the centroids of the clusters. Similar to OTSP, binary decision variables \(x_{ij} \in \{0,1\}\) indicate whether the direct route from location \(i\) to location \(j\) is included in the solution, where \(x_{ij} = 1\) signifies route inclusion. This means, an unconstrained VRP has a solution space of size $O(2^{n(n-1)})$. 

The original VRP of $12$ customer nodes and $1$ depot used as the example in this work, had $13 \times 12= 156$ binary decision variables representing edges between locations. To solve this problem on a quantum computer, one then needs $156$ logical qubits (assuming one-hot edge encoding). By using a clustered approach, we have redefined our example problem on $n=4$ nodes, i.e, our proposed solution architecture uses only $n(n-1)=12$ binary decision variables \(x_{ij}\), and this problem can be mapped on with only $12$ qubits. Thus, the hierarchical formulation maps each reduced subproblem to only 12 logical qubits, while the full 13-node edge-based formulation would require 156 logical qubits. This benefit comes at an extra cost of solving $3$ more $4$ node open loop traveling salesman problems \textit{sequentially} using the same qubits. In general, the qubit requirement for an $n-$node VRP is $O(n^2)$. In case of the clustered VRP, assuming each cluster contains $O(\sqrt{n})$ nodes, for sequential execution, the qubit requirements becomes $O(n)$, providing a square-root advantage. By increasing the number of clusters (i.e., number of sequential executions), the qubit-requirements can be decreased further.

The objective function that minimizes the inter-cluster travel distance is given as,   
\begin{equation}\label{cost_vrp}
H^C_{VRP}= \min \sum_{i=0}^{3} \sum_{j=0}^{3} w_{ij} x_{ij}, 
\end{equation} 

 where $w_{ij}$ are the elements of the distance matrix $W_C$. Similar to OTSP, the cost function in Eq.~\ref{cost_vrp} is also subjected to several constraints discussed as follows. We denote the depot as node $0$, and the centroids of the clusters as nodes $1$, $2$, and $3$.

\begin{itemize}
    \item \textit{Cluster Visit Constraints}: This constraint ensures that each customer cluster must be visited exactly once.
    
\begin{align*} 
\sum_{i \neq j} x_{ij} = 1, \quad & j \in \{1,2,3\},\\
& i \in \{0, 1, 2, 3\}-j. 
\end{align*}

\item \textit{Cluster Departure Constraints}: Each cluster must have exactly one departure: 

\begin{align*} 
\sum_{j \neq i} x_{ij} = 1, \quad & i \in \{1,2,3\},\\
&j \in \{0, 1, 2, 3\}-i. 
\end{align*} 

\item \textit{Vehicle Constraint}: The depot must have exactly 2 outgoing and 2 incoming edges (for 2 vehicles): 
\begin{align*}
\sum_{j} x_{0j} = 2, \quad &\sum_{i} x_{i0} = 2, \\
&i, j \in \{1, 2, 3\}-i. 
\end{align*} 

\end{itemize}

To address the constraints, we incorporate three penalty terms corresponding to each constraints as,

\begin{equation}
H_{CV}^{P} = \sum_{j=1}^{3} \left( 1 - \sum_{\substack{i=0 \\ i \ne j}}^{3} x_{ij} \right)^2
\end{equation}

\begin{equation}
H_{CD}^{P} = \sum_{i=1}^{3} \left( 1 - \sum_{\substack{j=0 \\ j \ne i}}^{3} x_{ij} \right)^2
\end{equation}

\begin{equation}
H_{VC} = \left( 2 - \sum_{j=1}^{3} x_{0j} \right)^2 + \left( 2 - \sum_{i=1}^{3} x_{i0} \right)^2
\end{equation}

The final VRP Hamiltonian incorporates objective and penalty terms as: 
\begin{equation} 
H_{\text{VRP}} = \sum_{i,j} w_{ij} x_{ij} + A (H^{P}_{CV}+H^{P}_{CD}+H^{P}_{VC}),
\end{equation} 
with a penalty parameter \(A \) chosen to be $100$. We then decompose this Hamiltonian in its corresponding QUBO and ISING form and proceed to solve the inter Cluster VRP with two different variants of the Quantum Approximate Optimization Algorithm, i.e., the standard QAOA \cite{Farhi2014}, and the multi-angle QAOA \cite{Herrman2022}. 

\noindent\textbf{Post-processing through Local bit-flip search}: For each OTSP and ICVRP instances, we employ standard and MA-QAOA variants with $50$ random initializations, and $10000$ measurement shots. Additionally, to ensure practical feasibility of routing solutions, we design and use a post-processing protocol based on exhaustive bit-flip local search over low Hamming-distance neighbors. The steps of the search is provided in Fig.~\ref{fig:methodology}(b). 

We found that the outcome of QAOA (both variants) circuits to solve the subproblems of the hierarchical VRP produces a probability distribution over a sub-space of the $2^{n_{q}}-$dimensional Hilbert space; $n_{q}$ is the number of qubits used in the quantum circuit. In the Fig.~\ref{fig:results}(a), (see Sec.~\ref{sec:results}), we provide histogram of the probabilities of basis vectors in the outcome of QAOA and MA-QAOA, averaged over all problem instances considered instances. 

The distribution clearly shows a few computational basis states appear with high probability in  both cases, with a tail of many states with exponentially small probabilities. It is comprehensible that depending on the number of measurements, only a fraction of the states in this superposition will be captured in an experimental scenario. 

We device a post-processing scheme that uses a realistic measurement budget to sample quantum states from this superposition of states created by QAOA. The sampled state then serves as a candidate to the VRP instance considered.  Following Fig.~\ref{fig:methodology}(b), the steps of the search is outlined below. Hereafter, we use 'QAOA' to represent both variants, unless otherwise mentioned. 

With a user-defined measurement budget $N_{m}$, the QAOA circuit is measured $N_{m}$ times, and a probability distribution of computational basis vectors (candidate bit-strings) is procured. Arranging the bit-strings in descending order of probability, we check if each candidate satisfies the constraints of problem instance. If the bit-string is feasible, it is accepted as a solution, and if not, we proceed to the next step. 

In the next-step, we exhaustively flip \textit{only} single and paired bits in the candidate bit-string, and check the feasibility of the flipped candidates. If at any stage we find a feasible solution, the post-processing algorithm stops.
If for any candidate bit-string, the exhaustive single and paired bit-flip searches fail (i.e., we find no feasible solutions), we move on to the next bit-string in the probability distribution, and continue the search as depicted in Fig.~\ref{fig:methodology}(b).

The design of the post-processing algorithm ensures polynomial complexity. In our simulations, we retain only $O(n_{q}^2)$ candidate bit-strings above a certain probability threshold. The single and paired bit-flip searches has complexity $O(n_q)$ and $O(n_q^2)$ respectively. Finally, since the verification of a proposed solution of an $NP$ problem is in $P$, the post-processing procedure is guaranteed to scale polynomially in $n_q$, provided the number of retained solutions is polynomial in $n_q$. Although the algorithm does not provide any deterministic guarantee of finding a solution, Fig.~\ref{fig:results} in Sec.~\ref{sec:results} shows that it significantly improves the probability of finding a feasible solution for both QAOA variants.

\noindent\textbf{Merging routing solutions}: After independently solving the intra-cluster routing problem (OTSP) and the inter-cluster routing problem, the two solutions are merged to obtain a complete feasible route for the original VRP instance. We show the merging technique in Fig.~\ref{fig:methodology}(c) with a representative dataset. 

In the figure, these intra-cluster OTSP paths are shown by the solid directed edges inside each cluster, while the triangular markers denote the cluster centroids.

Next, an inter-cluster VRP (ICVRP) is solved over the depot (yellow star) and the cluster centroids. This determines the order in which vehicles visit the clusters. The corresponding inter-cluster routes are shown by the dotted directed edges. For example, as shown in Fig.~\ref{fig:methodology}(c) vehicle $1$ follows
\[
\text{Depot} \rightarrow \text{Cluster 1 (blue)} \rightarrow \text{Depot},
\]
while vehicle $2$ follows
\[
\text{Depot} \rightarrow \text{Cluster 2 (green)} \rightarrow \text{Cluster 3 (pink)} \rightarrow \text{Depot}.
\]

To construct the final VRP route, each centroid visit in the inter-cluster solution is replaced by the corresponding intra-cluster OTSP path. Thus, when a vehicle is assigned to a cluster, it enters through the selected entry node, follows the OTSP path through all customers in that cluster, exits through the selected exit node, and then proceeds to the next cluster or returns to the depot according to the inter-cluster route.

The total cost of the merged VRP solution is computed as
\begin{equation}
C_{\text{total}}
=
C_{\text{ICVRP}}
+
\sum_{k=1}^{K} C^{(k)}_{\text{OTSP}},
\end{equation}
where $C^{(k)}_{\text{OTSP}}$ is the intra-cluster OTSP cost for cluster $k$, and $C_{\text{ICVRP}}$ is the inter-cluster routing cost over the depot and cluster centroids.

\section{Results}\label{sec:results} 
\begin{figure*}[!htbp]
    \centering

    \begin{minipage}[t]{0.298\textwidth}
        \centering
        \small{(a) QAOA Outcome}
        \vspace{0pt}

        \includegraphics[width=\linewidth]{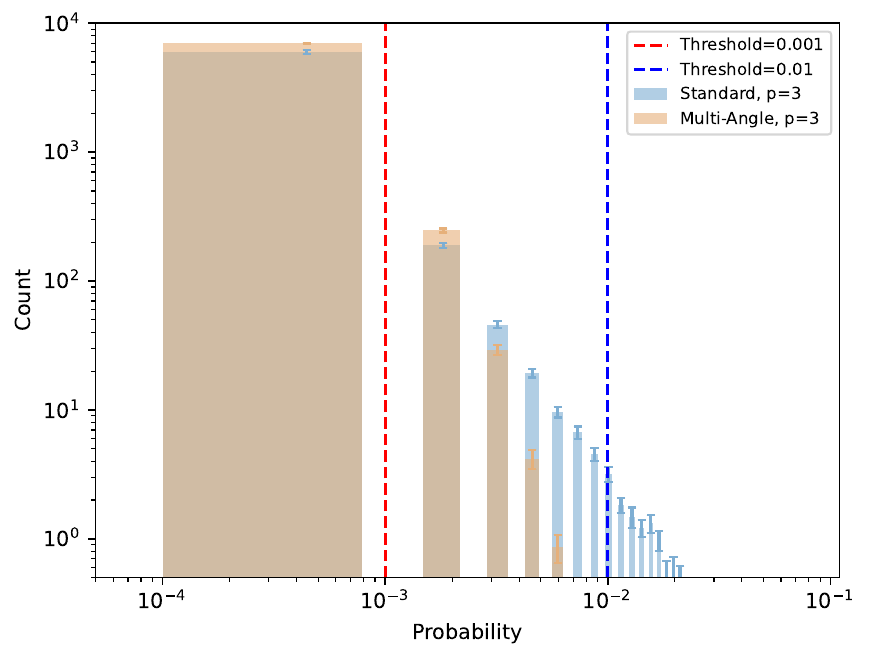}

        \vspace{3mm}

        \includegraphics[width=\linewidth]{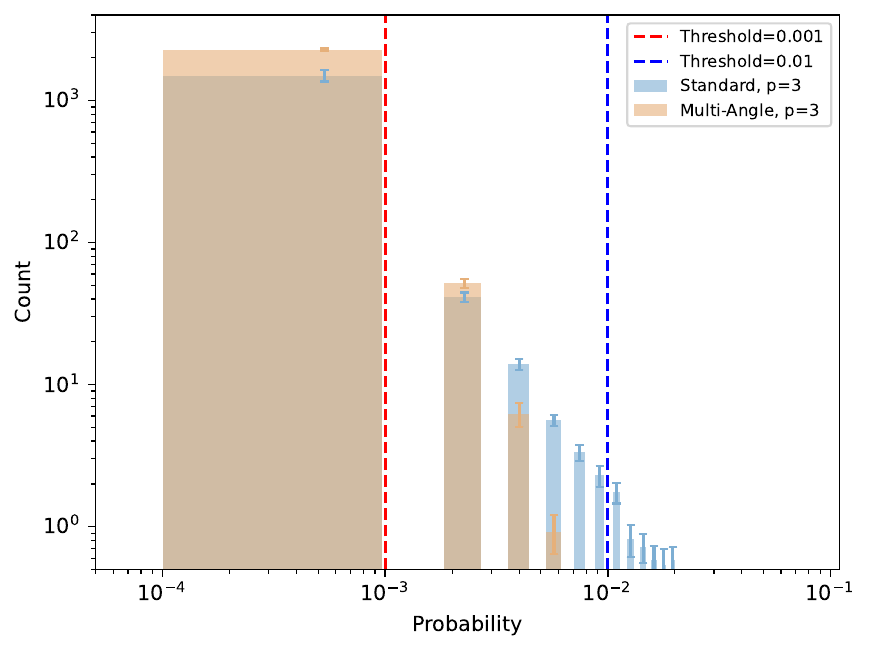}

        \vspace{1mm}
        
    \end{minipage}
    \hfill
    \begin{minipage}[t]{0.32\textwidth}
        \centering
        \small{(b) OTSP}
        \vspace{0pt}

        \includegraphics[
            width=0.715\linewidth,
            trim=0 0 150 0,
            clip
        ]{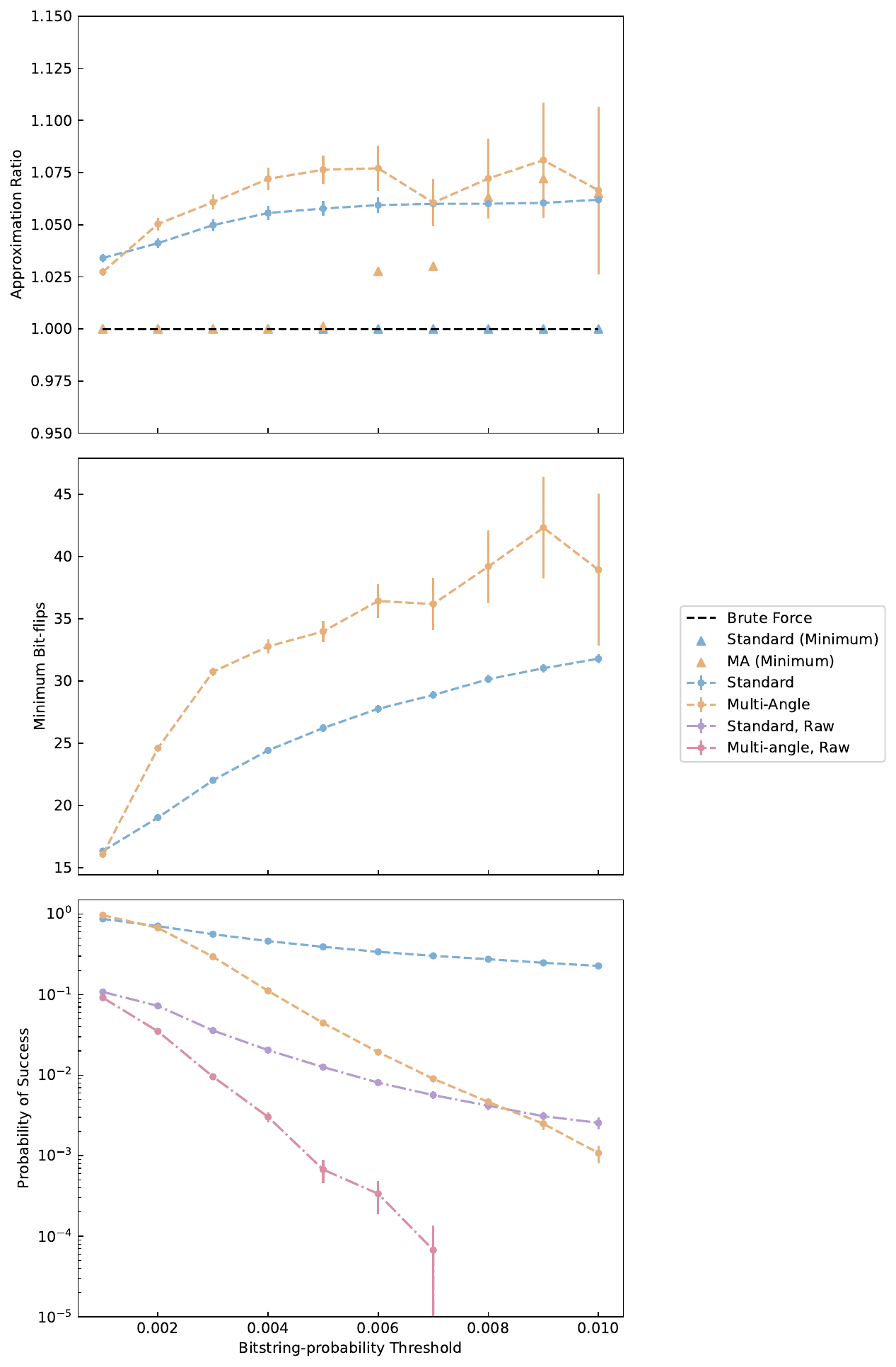}

        \vspace{1.3mm}
        
    \end{minipage}
    \hfill
    \begin{minipage}[t]{0.32\textwidth}
        \centering
        \small{(c) ICVRP}
        \vspace{0pt}

        \includegraphics[
            width=0.98\linewidth,
            trim=8 0 0 0,
            clip
        ]{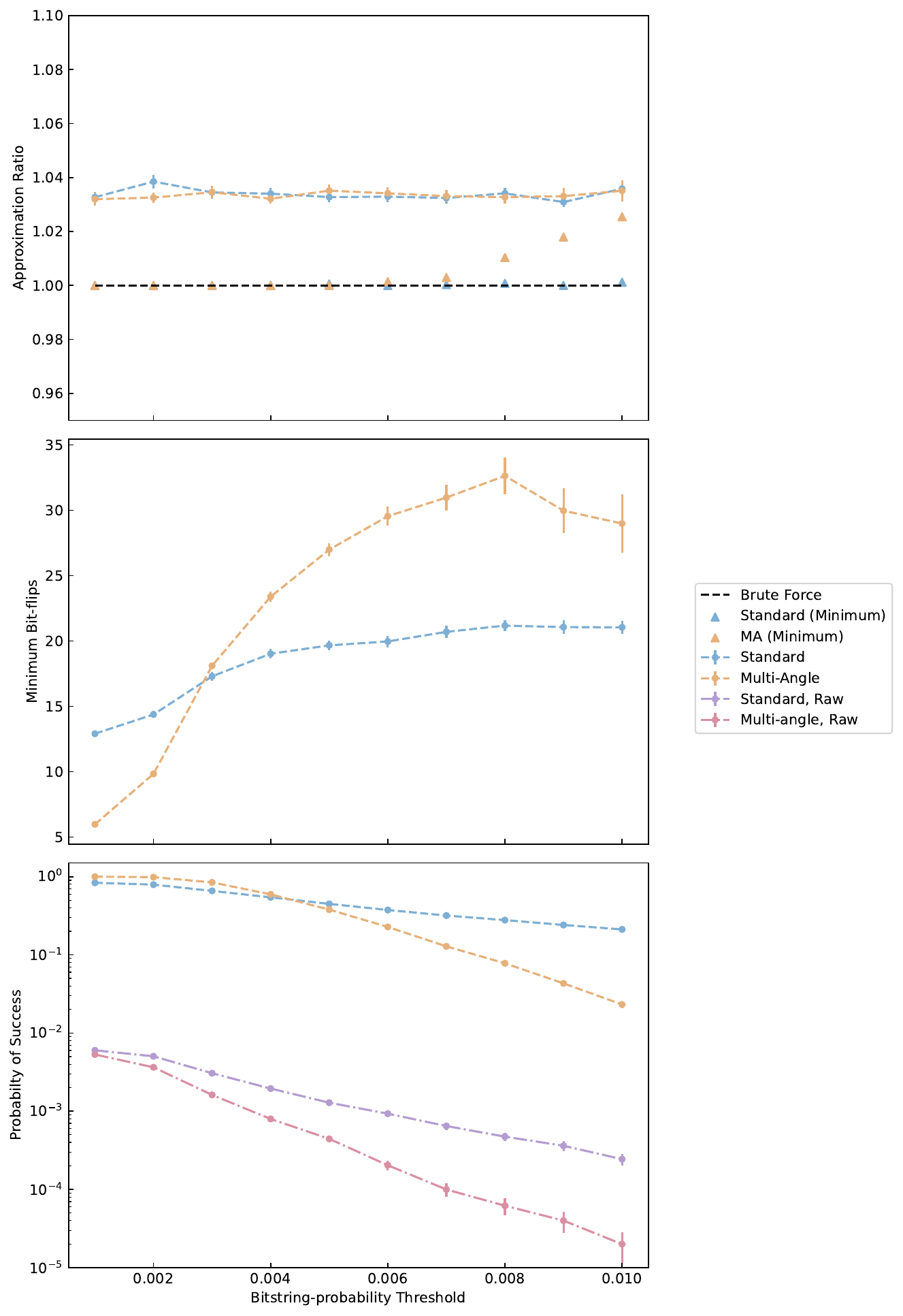}

        \vspace{1.1mm}
    \end{minipage}
    
    \caption{\textbf{(a)}: Histogram of the probability of computational basis vectors in the outcome of QAOA variants with depth $p=3$  on ICVRP (bottom) and OTSP (top) instances. The blue and yellow bars respectively correspond to the number of basis vectors (bit-strings) in different probability bins for standard and multi-angle (MA) QAOA outcomes. The red and blue dashed vertical lines mark probability thresholds of $0.001$ and $0.01$ respectively. These two values serve as the lower and upper ends of the threshold-range considered. Their placements in this figure is to provide an intuitive picture of the number of bit-strings retained at different choice of thresholds. \textbf{(b) [(c)]}: Benchmarking of OTSP [ICVRP] outcomes with increasing probability thresholds for the post-processing module. Bottom panel shows the average probability of success of the raw QAOA outcome (Standard: violet, MA: pink) and the same for the outcome passed through the post-processing module (Standard: blue, MA: yellow), computed with $50$ random initializations of QAOA per problem instance, and averaged over $300$ [$100$] problem instances. The middle panel shows the minimum number of bit-flips required by the post-processing module to find a feasible solution with increasing probability threshold (Standard: blue, MA: yellow), averaged over $300$ [$100$] problem instances, minimum taken over $50$ initiations of QAOA circuits. The top panel presents the approximation ratio against increasing probability threshold, computed against solutions found through a brute-force exhaustive search (black dashed line). The average approximation ratio for post-processed circuits is presented as blue (standard) and yellow (MA) dashed lines. Blue and yellow triangles represents the minimum approximation ratio per probability threshold value, minimum taken over $50$ independent runs of QAOA per problem instances. }
    \label{fig:results}
\end{figure*}

 We applied our hierarchical optimization framework to solve VRP with $2$ vehicles on $100$ randomly generated datasets. Each dataset contains $12$ customer nodes and $1$ fixed depot. For our simulations, we assume the depot to be the mean of the customer nodes. 
 
 For each VRP instance (dataset), we decomposed the set of $12$ customer locations to $3$ clusters using K-means clustering algorithm . Next, the intra-cluster routing for each cluster is implemented through solving OTSP for each dataset. We then compute inter-cluster routing by solving the $4$-node ICVRP instance per dataset. 
 
To solve each individual OTSP and ICVRP instance, we implemented both standard QAOA \cite{Farhi2014}, and MA-QAOA \cite{Herrman2022}, with $p=3$ layers. We have used IBM Qiskit to classically simulate both QAOA variants, with COBYLA as the underlying classical optimizer. Further, we ran $50$ randomly initialized QAOA circuit for each problem instance (for each dataset for ICVRP, and  for each cluster in every dataset for OTSP) for statistical rigor.  
 
 Next, for each of these $50$ runs, the probability distribution of the computational basis vectors $P(i)$ in the outcome of the quantum circuit was computed using $10000$ measurement shots. We then select a list of candidate bit-strings, ${\mathcal{L}}^i_{C}$ from $P(i)$, where each candidate has probabilities above a certain threshold $\tau$. This list is then passed on to the post-processing module presented in Sec.~\ref{sec:methods} and Fig.~\ref{fig:methodology}. To ensure that the number of candidate bit-strings to the post-processing module do not exceed $O(n^2)$, ($144 \text{ in our case}$), we imposed an additional condition on ${\mathcal{L}}^i_{C}$, such that, if $\lvert {\mathcal{L}}^i_{C} \rvert \geq 100$, only the top $100$ bit-strings with higher probabilities will be retained in ${\mathcal{L}}^i_{C}$. Provided this list, if the post-processing module outputs a feasible solution to the problem instance, we accept it, otherwise, the entire run is discarded.

We present our findings and benchmark the outcomes of the two QAOA variants in solving the individual OTSP and ICVRP instances  in Fig.~\ref{fig:results}. Fig.~\ref{fig:results}(a) shows the raw probability distributions of basis vectors in the measurement outcome of standard QAOA (blue) and MA-QAOA (yellow) circuits. The probability distribution for OTSP instances (top panel of Fig.~\ref{fig:results}(a)) is averaged over $50$ runs, each for $3$ clusters per 100 datasets, i.e., $15000$ circuits. As can be seen from the Figure, both the variants produces a probability distribution of a handful of states with higher probabilities, with a long tail of many states with very low probabilities. However, standard QAOA produces a tail with lesser number of states in comparison to MA-QAOA, and more number of states are found with higher probabilities. The red and blue dashed lines in the figure represents two values of probability thresholds, $\tau_{\rm {min}} = 0.001$ and $\tau_{\rm {max}}= 0.01$, as these are the extrema of the range of probability thresholds considered by us in selecting the candidate bit-strings for the post-processing module.

The bottom panel in Fig.~\ref{fig:results}(a) presents the probability distribution in the outcome of the QAOA circuits for ICVRP instances, averaged over $50$ runs per $100$ datasets, i.e., $5000$ circuits. The Figure portrays similar findings as OTSP, with bit-strings crossing the maxima of the threshold range for standard QAOA outcomes, while MA-QAOA produces a larger superposition of basis states with lower probabilities.

Fig.~\ref{fig:results}(b) presents the benchmarking results for the OTSP instances, for unprocessed and post-processed outcomes, with varying probability thresholds $\tau$. 
The bottom panel presents the probability of success $p(s)$ for a QAOA run at every threshold value. We present the success rates for both raw (violet: standard QAOA, pink: MA-QAOA), and post-processed outcomes (blue: standard QAOA, yellow: MA-QAOA). To clarify, here we define 'success' as finding any \textit{feasible} solution of the problem instance, and not the solution with lowest cost. Every datapoint in the Figure is the mean success probability in $50$ runs, averaged over $300$ problem instances ($3$ clusters per $100$ datasets). As is clear from the Figure, the 'raw' success probability, i.e., the probability of success for the bit-strings directly sampled from the QAOA circuits is significantly low compared to the post-processed outcome. Additionally, we note that the MA-QAOA performs worse than the standard variants for the entire range of probability thresholds. In comparison, the post-processing module significantly increases the probability of success, with both variants attaining $p(s)=1$ for lower threshold values. With increasing $\tau$, the success rates for MA-QAOA drops even with the post-processing module in-place, while, the standard QAOA maintains a high success rate throughout.

We present the minimum number of bit-flips required by the post-processing module to get a feasible solution of the corresponding OTSP instance in the middle panel of Fig.~\ref{fig:results}(b). The presented data is averaged over $300$ problem instances, while minimum is taken over the $50$ random initializations of QAOA variants (blue: standard, yellow: MA). We find that the standard QAOA requires less number of bit-flips to find a feasible solution for the problem considered, throughout the threshold range. 

Finally, we show the approximation ratio of the QAOA variants in the top panel of Fig.~\ref{fig:results}(b). The approximation ratio (AR) \cite{Farhi2014} is computed as, 
\begin{equation}\label{eq:AR}
    \text{AR} = \frac{\text{Total distance of the feasible path found}}{\text{Total distance of the optimal path}},
\end{equation}
where we calculate the optimal path through an exhaustive brute-force search over all possible $2^{12}$ bit-strings for benchmarking. As we are solving a minimization problem, $AR \geq 1$, and the equality represent the best solution found. We see that the on average, the AR (blue dashed line: standard, yellow dashed line: MA) remains within $1.05 \pm 0.05$ for both variants. The approximation ratio is computed after post-processing. Further, we note that the  minimum approximation ratio (blue triangles: standard, yellow triangles: MA) for standard QAOA remains $1$ for every value of the threshold $\tau$, whereas the MA-QAOA variant shows increase in approximation ratio for high values of $\tau$. The black dashed line represents the optimal value of AR.

Fig.~\ref{fig:results}(c) reports the benchmarking results of the ICVRP instances of the datasets considered. The bottom panel provides the probability of success $p(s)$ for the raw (violet: standard, pink: multi-angle),  and post-processed QAOA outcomes (blue: standard, yellow: multi-angle). We find that similar to the OTSP instances, the raw outcomes of the both variants of QAOA has very less probability of success, across the entire range of $\tau$. However, the local search post-processing module lifts $p(s)$ to a significantly high value for standard QAOA for all values of the bit-string probability thresholds. Further, MA-QAOA, aided with the post-processing module is able to attain high rates of success for $\tau \leq 0.006$. For the minimum number of bit flips (middle panel), we notice that the MA-QAOA requires less number of flips for low values of $\tau$, which increases at $\tau = 0.003$, whereas the standard variant requires roughly constant number of bit-flips across the entire range of $\tau$. However, the number of required bit-flips remain within $35$. We further find that the average approximation ratio $AR$, is approximately $1.04 \pm 0.001$ for both variants across the whole range of $\tau$, aided with the post-processing module. The minimum AR taken over the problem instances, similar to the OTSP, is $1$ for the standard QAOA for all values of $\tau$. For the MA-QAOA, the minimum grows around $\tau=0.007$, however, it maintains a lesser value than its OTSP counterpart.

Overall, we find that aided with the post-processing local bit-flip search, the standard QAOA performs better than the MA-QAOA for the individual OTSP and ICVRP instances. However, to note, as evidenced in Fig.~\ref{fig:results}(c), if the bit-string probability threshold is sufficiently low, MA-QAOA (with post-processing) can outperform the standard variant for ICVRP instances. We think, the complexity of the problem formulation in ICVRP favors the multi-angle optimization as done in MA-QAOA \cite{Shi2022}. Additionally, for ICVRP, when sampled with a low probability threshold, the average bit-flips required by the post-processing module to find a solution is significantly low $(\leq 5)$ for MA-QAOA implementation, (Fig.~\ref{fig:results}(c), middle panel). Based on this observation, we conjecture that with a higher value of QAOA layers $p$, the solution probability of MA-QAOA might increase for ICVRP instances.

\begin{figure*}[!htbp]
    \centering
    \begin{minipage}[c]{0.48\textwidth}
        \centering
         \small (a) Performance-Implementation
        \includegraphics[height=5.2cm, width=8cm, trim=8 0 0 0,
    clip]{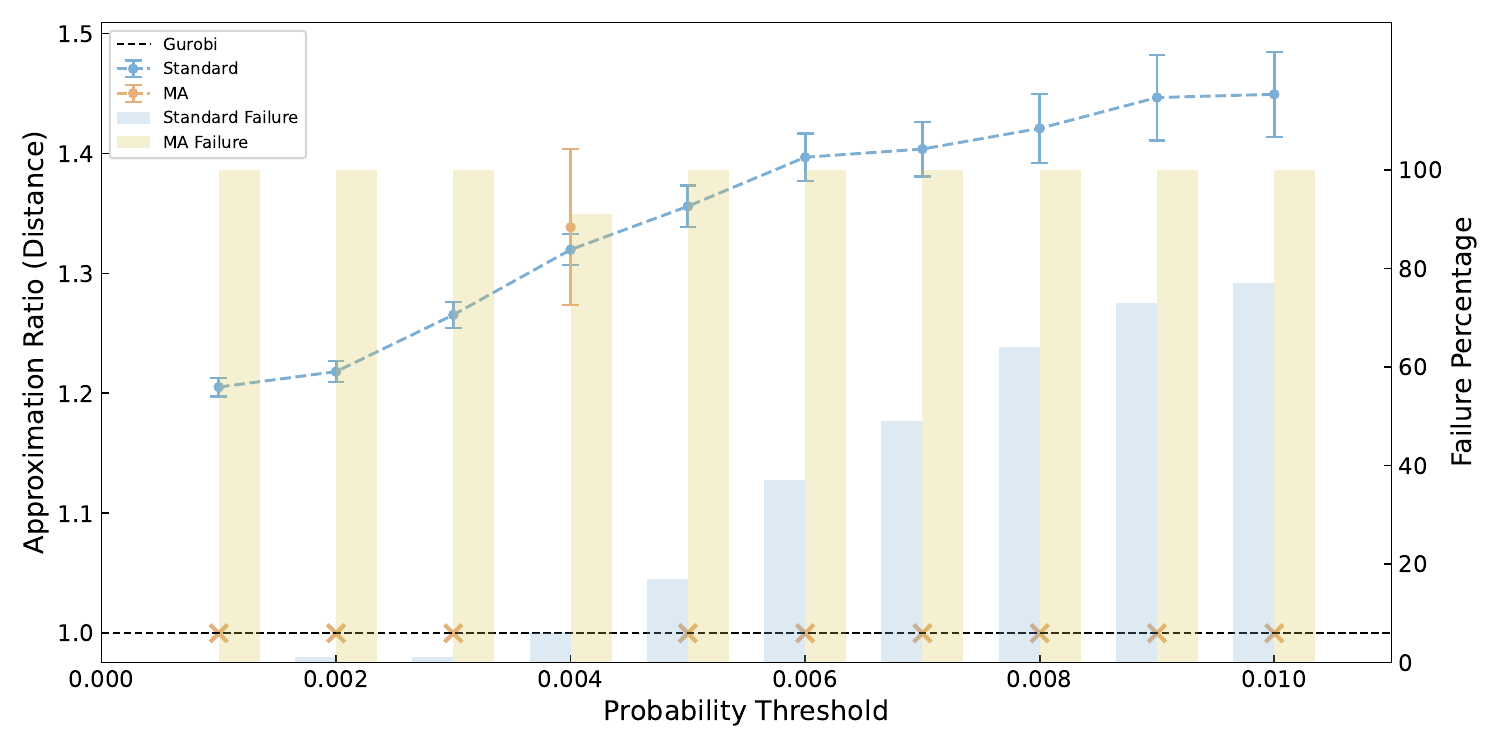}  
    \end{minipage}
    \hfill
    \begin{minipage}[c]{0.48\textwidth}
   
        \centering
        \small (b) Performance-Clusterability
        \includegraphics[height=5.2cm, trim=5 0 5 0,
    clip]{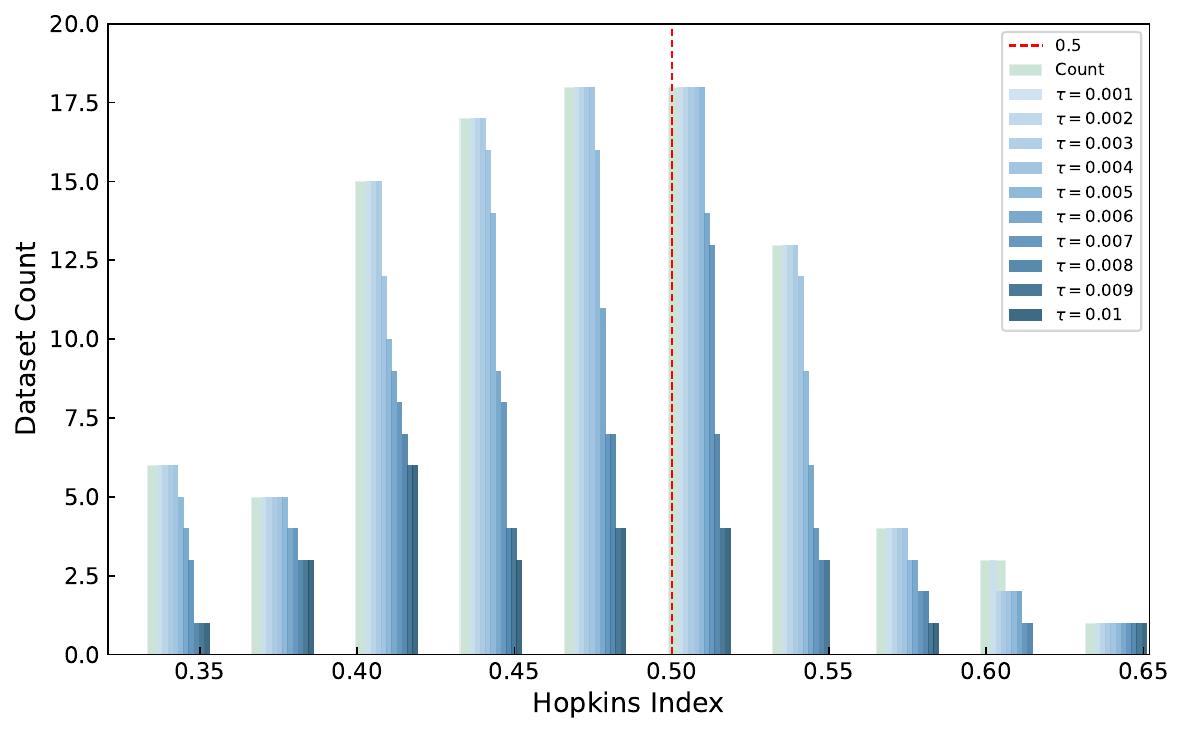}
       
    \end{minipage}
    \caption{Performance analysis of the hierarchical framework on the overall VRP. \textbf{(a)}: Comparison of the standard and MA-variants of the QAOA in hierarchically solving a $13$-node, $2$ vehicle VRP, against increasing probability threshold $\tau$ of the post-processing module. The blue dashed line represents the approximation ratio of the standard QAOA implementation, computed assuming solutions provided by a classical Gurobi optimizer as 'optimal'. The yellow circle represents the AR for MA-QAOA implementation. The black dashed line at $\text{AR}=1$ refers to the optimal Gurobi solutions, and yellow crosses denote no feasible solution found with MA-QAOA implementation. The blue and yellow bars respectively represents the failure percentage of standard and MA-QAOA implementations. \textbf{(b)}: Analysis of the performance of hierarchical algorithm (standard QAOA implementation, with different probability thresholds $\tau$ of the post-processing module), with respect to the clusterability (Hopkins index) of the datasets. The count of the datasets at certain clusterability index is depicted by the green bars. The blue bars denote the number of datasets for which the hierarchical algorithm find a feasible path. The progressively darker blue colors represent increasing values of $\tau$.}
    \label{fig:merged}
\end{figure*}
Once the solutions for the OTSP instances (for all clusters) as well as the ICVRP of a dataset are computed, we merge the routing paths of each cluster and compute the routing solution of the original $13$-node VRP, following the merging technique provided in Sec.~\ref{sec:methods}. To evaluate the effectiveness of our hierarchical approach, we compared the merged outcome of our algorithm with the outcome of the classical Gurobi optimizer \cite{gurobi} on the original $13$ node VRP for all datasets, since the VRP is written on $156$ binary variables, using an exhaustive search would have been inefficient and impractical. We present our results for the overall Vehicle Routing Problem in Fig.~\ref{fig:merged}(a). 

To merge the solutions, we have assumed that the each of the $50$ QAOA runs, for individual OTSP instances  corresponding to $3$ clusters of a single dataset and one ICVRP instances are connected. As an example, we consider dataset $d_{i}$. There are $3$ clusters, $C_1, C_2$ and $C_3$ for this dataset. In our simulations, we assume run $k$ of OTSP instances corresponding to clusters $C_1, C_2$ and $C_3$, and run $k$ of ICVRP instance for dataset $d_{i}$, completes a single run (run $k$) for the hierarchical framework, and we compare the solutions from only this run as possible merge candidate. 

This approach can provide us atmost $50$ merged candidate solutions for each dataset, each QAOA variant, and each probability threshold value. We compute all of the merged candidate solutions, if feasible (see Sec.~\ref{sec:methods}), and use them to compute the total distance corresponding to the original VRP. Then, the solution corresponding to the minimum distance is chosen as the \textit{final solution} of the hierarchical approach. Further, we compute the approximation ratio of this solution for each dataset, each QAOA variant, and each probability threshold value, assuming the solution path computed using the Gurobi optimizer to attain the lowest distance, i.e., 'optimal path' in Eq.~\ref{eq:AR}. 

The left-$Y$ axis in Fig.~\ref{fig:merged}(a) shows the approximation ratio of the both QAOA variants, averaged over $100$ datasets against the increasing probability threshold of the post-processing algorithm (blue dashed line: standard QAOA; yellow dot: MA-QAOA). The dashed black line at $\text{AR}=1$ line reflects the approximation ratio of Gurobi solutions. The right-$y$ axis provides the failure percentage of the hierarchical approach at each probability threshold value. The blue and yellow bars at each at each probability threshold $\tau$ respectively reflects the number of datasets (out of $100$) for which the standard and MA-QAOA implementations of the hierarchical approach did not find a feasible solution of the original $13$ node VRP. 

We find for standard QAOA, the hierarchical approach finds a feasible solution for $\approx 100 \%$ of the datasets for $\tau \leq 0.003$ and for $\approx 80 \%$ of the datasets  for $\tau \leq 0.005$, with an average approximation ratio within the range $1.2-1.4$. Above this threshold, the failure percentage of the standard QAOA implementation of the hierarchical algorithm increases, and at $\tau = 0.01$, a feasible solution is found for only $23$ datasets. The approximation ratio also increases with the increasing $\tau$, reflecting degradation of solution quality. Interestingly, the MA-QAOA implementations found a solution for the overall VRP at only $\tau = 0.004$, with an $\text{AR} = 1.338 \pm 0.065$, and a success probability of only $9 \%$. 

While standard QAOA with post-processing provides high average success rate and minimum approximation ratio $1$ in solving single instances of OTSP and ICVRP across the entire range of $\tau$, our assumption of initializing QAOA to solve the entire problem with a single initiation index (a single 'run' for four underlying sub-problems; see above) leads to the degradation of results at higher threshold values.The success rate will increase if one is allowed to treat the each run of the sub-problems individually, and then select the best combination. However, with $50$ runs per subproblem, the total number of possible combinations is huge, leading us to use the current convention. The failure of MA-QAOA is closely related, and also tied to the fact that MA-QAOA implementations work worse in solving OTSP instances (Fig.~\ref{fig:results}(b)), leading to no feasible solutions for the overall problem in most of the cases. 

 Finally, we investigated the effect of clusterability of a dataset on the hierarchical algorithm. Using Hopkins statistics (Hopkins index) $H$ \cite{lawson1990hopkins} to analyze the clusterability of the datasets, we identify the clusterability of the $100$ randomly generated datasets in Fig.~\ref{fig:merged}(b)). Among the generated datasets, $68$ have $H < 0.5$, and $32$ have $H > 0.5$. The value of $H = 0.5$ (red dashed line in Fig.~\ref{fig:merged}(b)) serves as a threshold of the clusterability of the datasets, as $H \leq 0.5$ indicates no meaningful clusters exists, and $H > 0.5$ and above implies there is a high tendency of clustering. We then compare the performance of the hierarchical optimization on both strongly and weakly clusterable customer nodes in Fig.~\ref{fig:merged}(b). 
 
 Fig.~\ref{fig:merged}(b) shows a histogram of the datasets, divided in $10$ bins with respect to their $H$, and the green bar, placed at the bin-mean, indicates the number of the datasets in the corresponding bin. The gradually darker blue bars then indicate the successful counts of the hierarchical standard QAOA implementations for the datasets in that corresponding bin, with the color gradients representing increasing $\tau$ values. Here 'success' indicates finding a feasible solution for the overall VRP. We find that clusterability has negligible impact on the hierarchical algorithm, as the height of the blue bars is consistent with dataset counts, and the total failure percentage, as shown in Fig.~\ref{fig:merged}(a). As the success rate is $0$ for most of the threshold values for the MA-QAOA implementations (see Fig.~\ref{fig:merged}(a)), we did not include this variant in this analysis.

\section{Conclusion}\label{sec:conclusion}
In this work, we demonstrate that the Quantum Approximate Optimization Algorithm (QAOA) can be useful to hierarchically solve large-scale Vehicle Routing Problems. The hierarchical decomposition clusters the customer nodes into smaller sub-groups based on their distances, and presents the original problem as a Vehicle Routing Problem between the depot and the centroid of the customer cluster. We present a rigorous mathematical formulation of the hierarchical decomposition, proposing to solve open loop traveling salesman problem (OTSP) within the clusters, and finally solving a smaller inter-cluster Vehicle Routing Problem (ICVRP), as presented above. 

We benchmark our approach on $100$ synthetically generated datasets with $12$ customer nodes, $1$ depot, and $2$ vehicles. We divide each set of customer nodes into $3$ clusters of $4$ nodes, and solve $4$ problems (3 OTSP, $1$ ICVRP) per dataset using two variants of  QAOA, the standard variant, as proposed in \cite{Farhi2014}, and the multi-angle  QAOA (MA-QAOA) variant \cite{Herrman2022, Shi2022}. Further, we propose a $1$ and $2$ local bit-flip search post processing algorithm that samples solutions from the QAOA output based on a probability threshold $\tau$. We show that, this algorithm aids the QAOA variants to achieve significantly higher solution probability and approximation ratio. For each small subproblems, we benchmark the $12$ qubit QAOA solutions against exhaustive search. 

Further, we implement a merging technique to solve the original VRP, and benchmark the corresponding results against classical Gurobi optimizer, that solve the $156$ variable problem directly. We show that standard QAOA, with $p=3$, and aided with a simplistic post-processing algorithm finds a solution of the original VRP within approximation ratio $1.2$, and $100 \%$ success rates in the best cases. MA-QAOA, while performs well for ICVRP instances when allowed a sufficiently low threshold for the post-processing module, fails to reliably generate a solution for the original VRP in most cases. Additionally, we show that clusterability of a customer dataset has no significant impact on the solution quality of the standard QAOA implementation. 

The hierarchical decomposition can be used recursively, i.e., the OTSP within the clusters can be further clustered and solved using the same technique as presented in this work. As is well known, to solve an $n$ node VRP with QAOA, $O(n^2)$ qubits are required. At current age of quantum computing, where the modern quantum computers struggle to achieve $\approx 10$ logical qubits, our results indicate that hierarchical decomposition can substantially reduce qubit requirements for QAOA-based routing heuristics, and may provide a useful direction for applying near-term quantum algorithms to larger structured optimization problems.

\section{Data Availability}\label{sec:data}
Presently, the data and the code are available from the authors on reasonable request. We plan to open-source the code at a later time. 

\bibliographystyle{apsrev4-2}

\end{document}